\documentclass[twocolumn,prl,superscriptaddress,footinbib,longbibliography]{revtex4-1}
\pdfoutput=1
\usepackage[utf8]{inputenc}
\usepackage{amsmath}
\usepackage{amssymb}
\usepackage{graphicx}
\usepackage{esint}
\usepackage[lBrack,rBrack,llparenthesis,rrparenthesis]{stmaryrd}

\usepackage{ifpdf}
\ifpdf
  \DeclareGraphicsExtensions{.pdf,.png,.jpg}
\else
  \DeclareGraphicsExtensions{.eps}
\fi

\usepackage{graphicx}
\graphicspath{{fig/}{./}{work/gatsby/result_kps/}{work/gatsby/time/}}

\newcommand{\ec}[1]{\mathbb{E}\left[#1\right]}
\newcommand{\vc}[1]{{\mathbb{V}}\left[#1\right]}
\newcommand{\dc}[1]{{\mathbb{D}}\left[#1\right]}

\newcommand{\dcc}[2]{{\mathbb{D}_{#1}}\left[#2\right]}

\newcommand{\eca}[1]{\mathbb{E}^\mathrm{a}\left[#1\right]}
\newcommand{\eco}[1]{\mathbb{E}^\mathrm{o}\left[#1\right]}

\newcommand{\D}{\mathbf{D}}
\newcommand{\bmu}{\bm{\epsilon}}
\newcommand{\dis}{\mathbf{d}}
\newcommand{\x}{\mathbf{r}}
\newcommand{\m}{\mathbf{e}}
\newcommand{\del}[3]{ #1_{ #2 \shortrightarrow #3} }

\newcommand{\delx}[2] {\x_{ #1 \shortrightarrow #2} }

\makeatletter

\newcommand{\green}[1] {\textcolor{black}{#1}}
\newcommand{\blue}[1] {\textcolor{black}{#1}}
\DeclareFontFamily{OT1}{pzc}{}
\DeclareFontShape{OT1}{pzc}{m}{it}{<-> s * [1.05] pzcmi7t}{}
\DeclareMathAlphabet{\mathpzc}{OT1}{pzc}{m}{it}

\def\@dotsep{4.5}

\usepackage{epsf}
\usepackage{graphics}
\usepackage{rotating}
\usepackage{bm}
\usepackage{enumerate}
\usepackage{color}

\makeatother

\begin{document}

\title{Mass transport computations via correlation splitting and a law of total diffusion}

\author{Manuel Ath\`enes}
\affiliation{Universit\'e Paris-Saclay, CEA, Service de Recherches de Métallurgie Physique, 91191, Gif-sur-Yvette, France}

\author{Gilles Adjanor}
\affiliation{Groupe M\'etallurgie, MMC, EDF, Les Renardi\`eres, 77818 Moret-sur-Loing, France}

\author{J\'er\^ome Creuze}
\affiliation{Universit\'e Paris-Saclay, ICMMO, CNRS UMR 8182, F91405 Orsay Cedex, France}







\begin{abstract} 
Directly computing mass transport coefficients in stochastic models requires integrating over time the equilibrium correlations between atomic displacements. Here, we show how to accelerate the computations via \green{correlation splitting and conditioning, which statistically amounts to estimating the mass transport coefficients} through a law of total diffusion. We illustrate the approach with kinetic path sampling simulations of atomic diffusion in a \green{random alloy model} in which percolating solute clusters trap the mediating vacancy. There, Green functions serve to generate first-passage paths escaping the traps and to propagate the long-time dynamics. When they also serve to estimate mean-squared displacements via conditioning, colossal reductions of statistical \green{errors} are achieved.
\end{abstract}

\maketitle

\blue{\section{Introduction}}
Mass transport is the natural phenomenon that governs both the time evolution of thermodynamic systems towards equilibrium and the diffusion of chemical species at equilibrium. Its understanding is fundamentally important and challenging in many engineering applications, ranging from biophysics to materials science. Equilibrium correlations between particle displacements play a crucial role because they allow characterizing transport coefficients close to equilibrium. To give a typical example, chemical currents in linear response theory~\cite{chandler:1987,evans:2007} are expressed as products of diffusion matrices and negative gradients of chemical concentrations considered as small thermodynamic forces. Kinetic Monte Carlo (kMC)~\cite{bortz:1975} provides a direct method to measure  transport coefficients in multi-dimensional systems governed by master equations. However, the method is hindered by the low occurrence of important events. A considerable amount of computations is often needed to collect sufficient statistics in many systems of interest~\cite{trochet:2017,trochet:2019}, including basic models of defect migration through homogeneous media~\cite{swinburne_perez:2020}. Besides, low-complexity models are amenable to non-stochastic master-equation approaches~\cite{koiwa:1984,nastar:2014,bocquet:2014,vaks:2016} that may yield accurate predictions at much lower cost. These methods involve integrating fundamental matrices~\cite{koiwa:1984,bocquet:2014,schuler:2020} or equivalently evaluating lattice Green functions~\cite{trinkle:2017,agarval:2017} that either minimize a variational problem~\cite{arita:2017,arita:2018,trinkle:2018} or satisfy a Poisson equation~\cite{duflo:1997,veretennikov:2017,delmas:2009,athenes:2018}. The dimension of the linear systems to solve and the viability of approximations made to handle exponentially increasing numbers of configurations limit the  applicability of non-stochastic approaches. 

Here, we develop a \green{correlation splitting} and conditioning scheme aiming to facilitating the computations of mass transport coefficients. The approach alleviates the aforementioned issues: scarcity of harvested important events and combinatorial explosion in numerical algebra. It furthermore leads us to formulate a law of total diffusion (LTD) that relates to the laws of total expectation (LTE) and variance (LTV) considered so far to compute thermodynamic expectations and their statistical variance via conditioning~\cite{frenkel:2004,frenkel:2006,athenes:2007,delmas:2009,athenes:2010,adjanor:2011,cao:2014,athenes:2017,frenkel:2017,athenes:2018}. We illustrate the approach on a \green{random} alloy model~\cite{trinkle:2018} exhibiting dynamical trapping and percolation by estimating the diffusion matrix, denoted below by $\dc{\mathbf{d}}$ and defined as half the asymptotic variance of the vector $\mathbf{d}$ of chemical displacements. 

\blue{\section{Diffusion matrix for reversible Markov chains}}

To monitor the displacements of each of the $c$ chemical species along each of the $s$ space dimensions, any state $\chi$ is mapped onto a descriptor $\mathbf{r}\in\mathbb{R}^{sc}$. Component $r_{i}$ represents here the coordinate sum of all $\alpha$-type atoms in space direction $a$, the descriptor index being encoded as $i=a+s(\alpha-1)$, where $a\in\mathbb{N}^\star_s$ and $\alpha\in\mathbb{N}^\star_c$. Then, displacement $\dis(\chi,\chi')$ from state $\chi$ to state $\chi'$ is computed using the minimum image representation of $\mathbf{r}'-\mathbf{r}$~\footnote{$\dis(\chi,\chi^\prime)$ equals $\x^\prime-\x + \arg\min \protect\|\x^\prime-\x + \mathbf{m} \protect\|$, where the components $m_i$ of argument $\mathbf{m}$ run over the multiples of the simulation supercell periods.}. 
The variance matrix of vector $\x$ over a Markov chain reads $\vc{\x}=\ec{\x^2}-\ec{\x}^2$, with $\x^2$ standing for tensorial square $\x\otimes\x$ and $\ec{\cdot}$ denoting the expectation \blue{operator}. The successive states of the chain are denoted by $\chi_h$ with $h\in \mathbb{N}$ or $\mathbb{Z}$. The displacement vector after $\ell$ transitions, $\sum_{h=0}^{\ell-1} \dis\left( \chi_h,\chi_{h+1} \right)$, is simply written $\x_{0 \shortrightarrow \ell}$ in the following. The expected square displacement divided by twice the elapsed time is written
\begin{equation}
\dcc{\ell}{\mathbf{d}} = \frac{1}{2} \frac{\ec{\delx{0}{\ell}\otimes \delx{0}{\ell}}  } {\ec{\del{t}{0}{\ell}} } = \frac{1}{2} \frac{\vc{\delx{0}{\ell}}} {\ec{\del{t}{0}{\ell}} }.   \label{def1}
\end{equation}
The asymptotic limit yields the diffusion matrix, i.e., $\dc{\dis} = \lim_{\ell \shortrightarrow \infty}\dcc{\ell}{\mathbf{d}} $. The variance amounts to squaring here because the expected displacement $\ec{\x_{0 \shortrightarrow \ell}} $ is zero at equilibrium. The Markov chain obeys detailed balance with respect to an equilibrium stationary distribution $\rho^\mathrm{eq}_\chi$. Fulfillment of this strong condition entails the invariance of expectations under arbitrary translation of chain indexes and under chain reversal. Expected autocorrelations are thus invariant after interchange of displacements: $\ec{ \x_{0 \shortrightarrow 1} \otimes \x_{h \shortrightarrow h+1}}$ is successively equal to $\ec{\x_{-1-h\shortrightarrow -h} \otimes  \x_{-1 \rightarrow 0}} $, $\ec{\x_{h+1\shortrightarrow h} \otimes  \x_{1 \shortrightarrow 0}}$ and eventually $\ec{  \x_{h\shortrightarrow h+1} \otimes  \x_{0 \shortrightarrow 1}}$, displacements being anti-symmetrical under chain reversal: $\x_{h+1 \shortrightarrow h}=-\x_{h \shortrightarrow h+1}$, $\forall h$. Hence, $\dc{\mathbf{d}}$ is symmetric nonnegative. Translational invariance also entails that the expected elapsed time is $\ell$ multiplied by the mean elapsed time before the next-event $\bar{\tau} = \ec{t_{0\shortrightarrow 1}}$. The autocorrelation symmetry properties enable us to split the diffusion matrix into two parts: 
\begin{align}\label{eq:Dsum}
 \dc{\dis} = \dcc{1}{\dis} + \frac{1}{\bar{\tau}}\sum\nolimits_{h=1}^{\infty} \ec {\x_{0 \shortrightarrow 1} \otimes \x_{h \shortrightarrow h + 1}}.  
\end{align}
For uncorrelated chains, the \emph{uncorrelated} part $\dcc{1}{\dis}=\ec{\delx{0}{1}\otimes\delx{0}{1}}/\left[ 2\bar{\tau} \right]$ contributes to diffusion exclusively, since the \emph{correlated} part encompassing the summation vanishes. For low-dimensional spaces,  $\bar{\tau}$  and $\dcc{1}{\dis}$ can be readily evaluated from the knowledge of $\rho_\chi^\mathrm{eq}$ via conditioning on the states. Let $\ec{f(\chi_k,\chi_{k+1})|\chi_0}$ denote the conditional expectation of function $f(\chi_k,\chi_{k+1})$ given $\chi_0$ and integer $k\geq 0$. The mean first-passage time (MFPT) from $\chi_0$ then writes $\tau_{0}=\ec{t_{0 \shortrightarrow 1}|\chi_0}$ and its mean yields $\bar{\tau} = \ec{\tau_0}$. Term $\ec{\delx{0}{1}\otimes\delx{0}{1}}$ can be similarly evaluated, yielding the uncorrelated part. Evaluating the correlated part is however more difficult because it requires formulating and solving a Poisson equation. For this purpose, define the mean displacement from $\chi_1$ as 
\begin{align}
\m(\chi_1)=\ec {\x_{1\shortrightarrow 2}|\chi_1} =-\ec {\x_{0\shortrightarrow 1}|\chi_1}  \label{eq:reversibility_property}, 
\end{align}
and the associated relaxation vector $\bmu(\chi_1) = \sum\nolimits_{h=1}^{+\infty}\ec{\m(\chi_{h})|\chi_1}$ whose knowledge will give access to the correlated part of the diffusion matrix. The relaxation vector is a particular solution of the following discrete Poisson equation
\begin{align} 
\bmu(\chi_1) = \ec{\bmu(\chi_2)|\chi_1} + \m(\chi_1) \label{eq:poisson}
\end{align}
where $\chi_1$ runs over the state space. The solution $\bmu$ is fully determined by additionally imposing $\ec{\bmu}=\ec{\m}=0$. It characterizes the expected correlations:  
\begin{align} 
\ec {\x_{0\shortrightarrow 1} \otimes \bmu(\chi_1)} &= \sum\nolimits_{h=1}^\infty \ec{\ec {\x_{0\shortrightarrow 1} \otimes \x_{h\shortrightarrow h+1}|\chi_0,\chi_1}},\nonumber\\
 &= \sum\nolimits_{h=1}^\infty \ec{\x_{0\shortrightarrow 1} \otimes \x_{h\shortrightarrow h+1}}. 
\label{eq:eps} 
\end{align}
Combining~\eqref{eq:Dsum} and~\eqref{eq:eps} then yields the diffusion matrix:
\begin{align}
\begin{aligned}
\dc{\delx{0}{1}} &= \dcc{1}{\delx{0}{1}} + \frac{1}{\bar{\tau}}\ec {\x_{0\shortrightarrow 1} \otimes \bmu(\chi_1)}  \\
                 &= \frac{1}{2\bar{\tau}} \ec{ \left( \x_{0\shortrightarrow 1} + \bmu(\chi_1)\right)^2 -\bmu^2 }, 
\end{aligned}\label{eq:Dsym}
\end{align}
where first equality is similar to Green-Kubo formulae~\cite{arita:2017,arita:2018} and symmetry is formally recovered in second equality. Expressions in~\eqref{eq:Dsym} are the cornerstone of variational approaches to mass transport~\cite{arita:2017,arita:2018,trinkle:2018} and serve here the purpose of  improving kMC measurements via \green{correlation splitting} and conditioning. 

\blue{\section{Correlation splitting and conditioning}}

We proceed by first writing the LTV for displacement $\delx{1}{2}$ with conditioning on $\chi_1$ and rescaling with $2\bar{\tau}$,
\begin{align}\label{eq:ltv}
\dcc{1}{\delx{1}{2}}  = \ec{\dcc{1}{\delx{1}{2} |\chi_1}} + \dcc{1}{\ec{\delx{1}{2} | \chi_1}}, 
\end{align}
where $\dcc{1}{\delx{1}{2} |\chi_1}=\vc{\delx{1}{2} |\chi_1}/\left[2 \bar{\tau}\right]$ and the conditional variance is  
$\vc {\delx{1}{2} |\chi_1} =\ec{(\x_{1\shortrightarrow 2})^2|\chi_1}-\m(\chi_1)^2$. 
Further \green{splitting} the diffusion matrix~\eqref{eq:Dsym} is then obtained through: (i) inserting the LTV~\eqref{eq:ltv} in~\eqref{eq:Dsym} and conditioning the expectation of $\x_{0\shortrightarrow 1} \otimes \bmu(\chi_1)$ on $\chi_1$; (ii) inserting the reversibility property~\eqref{eq:reversibility_property}; (iii) plugging the Poisson equation~\eqref{eq:poisson}; (iv) simplifying the plugged expectation $\ec{\ec{\m(\chi_1)\otimes \bmu(\chi_2)|\chi_1}}$ into $\ec{\m(\chi_1)\otimes \bmu(\chi_2)}$ using the LTE; (v) regrouping like quantities $\dcc{1}{\m}=\ec{\m^2}/\left[2\bar{\tau}\right]$;  (vi) identifying the diffusion matrix of the conditionally expected displacements,   
\begin{align}
\begin{aligned}
 \dc{\m(\chi_1)} &=  \dcc{1}{\m(\chi_1)} +  \frac{1}{\bar{\tau}}\ec{ \m(\chi_1)\otimes \bmu(\chi_2)} \\
 &= \frac{1}{2\bar{\tau}}\ec{\left( \m(\chi_1) + \bmu(\chi_2)\right)^2-\bmu^2}.
 \end{aligned} \label{eq:extra}
\end{align}
by analogy to expressions of Eq.~\eqref{eq:Dsym}. Both contributions~\eqref{eq:Dsym} and~\eqref{eq:extra} are measurable by kMC simulations. The resulting splitting yields the LTD:  
\begin{align}
 \dc{\del{\x}{1}{2}}  &=\ec{\dcc{1}{\del{\x}{1}{2} |\chi_1 }} - \dc{\ec{\del{\x}{1}{2} | \chi_1}}, \label{eq:main} 
\end{align}
in which the diffusion matrix is expressed as the difference between two symmetric nonnegative contributions, the \emph{intra}- and \emph{extra}-correlated diffusion matrices, respectively. Intra-correlations involve consecutive displacements exclusively, since $\ec{\dcc{1}{\delx{1}{2}|\chi_1}}=\dcc{2}{\delx{0}{1}}$. \green{The remaining correlations contribute to the  extra-correlated part, 
$\dc{\m(\chi_0)} = -\ec{\x_{0 \shortrightarrow 1} \otimes\left(  \x_{1 \shortrightarrow 2} +2\sum\nolimits_{h=2}^{\infty} \x_{h \shortrightarrow h+1}\right)}/[2\bar{\tau}]$. 
The former} equality arises as the particular case $\ell=2$ of a more general relationship, 
\begin{align}
\dcc{\ell}{\dis}  =\ec{\dcc{1}{\dis |\chi}} + \tfrac{1}{\ell} \dcc{1}{\ec{\dis | \chi}} - \tfrac{\ell-1}{\ell} \dcc{\ell-1}{\ec{\dis|\chi}},  \label{eq:LTC}
\end{align}
derived from the reversibility property~\eqref{eq:reversibility_property} and the law of total covariance~\cite{noteSM}, where $\chi\equiv\chi_1$ and $\dis\equiv\delx{0}{1}$. Relationship~\eqref{eq:LTC} bridges between the LTV at the lower extremity and the LTD obtained for $\ell\rightarrow \infty$. \green{The LTD entails that the diffusion matrix can be estimated from mean local quantities, by plugging the expected displacements and their conditional variances given the visited states into its extra- and intra-correlated parts.} \blue{Besides, the LTD and bridging law are meaningful for reversible Markov chains and for any stochastic variable that is antisymmetric under chain reversal. } 
Laws~\eqref{eq:ltv} and~\eqref{eq:main} also yield a Löwner partial ordering:  
\begin{align*}
 \dc{\dis} \leq \ec{\dcc{1}{\dis|\chi}} \leq \dcc{1}{\dis},  
\end{align*}
entailing that $\dc{\dis}$ is better approximated by its intra-correlated part $\dcc{2}{\dis}$ than its uncorrelated part $\dcc{1}{\dis}$. Further increasing $\ell$ provides a decaying sequence $\left( \dcc{\ell}{\dis}\right)_{\ell \geq 1}$ of upper-bound approximates of $\dc{\dis}$~\cite{trinkle:2018}. We then estimate $\dc{\dis}$ via $\dcc{\ell}{\dis}$ in~\eqref{eq:LTC} over a sample of $l$ trajectories of $L=\ell_\mathrm{max}$ steps each. For statistical errors to be small, $l$ must be large enough, standard deviations decaying as $1/\sqrt{l}$~\cite{caflisch:1998}. Letting $\chi_{i,h}$ denote state $h$ of trajectory $i$ and  $\mathbf{V}(\chi_{i,h})$ stand for $\vc{\delx{i,h}{h+1}|\chi_{i,h}}$, the LTD-based conditioned estimator of $\dcc{\ell}{\dis}$ writes
\begin{align}\label{eq:condestimator}
\D_{l,\ell}^\mathrm{cnd}  & = \frac{1}{2\widehat{\tau}} \cdot \frac{1}{lL}\sum\nolimits_{i=1}^l \sum\nolimits_{h=0}^{L-1} \left[\mathbf{V}(\chi_{i,h})+\tfrac{1}{\ell}\mathbf{e}(\chi_{i,h})^2\right] \nonumber \\
&-  \frac{1}{2\widehat{\tau}} \cdot \frac{1}{l\ell}\sum\nolimits_{i=1}^l \left(\sum\nolimits_{h=1}^{\ell-1} \m(\chi_{i,h}) \right)^2 , 
\end{align}
where $\widehat{\tau}$ is an estimate of $\bar{\tau}$ obtained via LTE-based conditioning, i.e., by averaging the MFPTs~\cite{athenes:1997,puchala:2010}: 
\begin{align} \label{eq:time_conditioning}
 \widehat{\tau} = \frac{1}{l L}\sum\nolimits_{i=1}^l \sum\nolimits_{h=0}^{L-1} \ec{t_{i,h\shortrightarrow h+1} | \chi_{i,h} }.
\end{align}
Note that information about the $lL$ gathered states is included in the estimation of static expectations in~\eqref{eq:condestimator} and~\eqref{eq:time_conditioning}. Conditioned estimator in~\eqref{eq:time_conditioning} has a statistical variance that is lower than that of the plain estimator $\sum_{i=1}^l t_{i,0\shortrightarrow \ell}$, a property guaranteed by the LTV~\cite{frenkel:2006,athenes:2017,athenes:2018}. The conditioning over time is traditionally done in the standard estimator of $\dcc{\ell}{\dis}$:
\begin{align}\label{eq:stdestimator}
\D^\mathrm{std}_{l,\ell} = \frac{1}{2\widehat{\tau}}\cdot \frac{1}{l\ell} \sum\nolimits_{i=1}^l \x_{i,0\shortrightarrow \ell} \otimes \x_{i,0\shortrightarrow \ell}. 
\end{align} 
\green{While estimators~\eqref{eq:condestimator} and~\eqref{eq:stdestimator} are both valid, their statistical variances usually differ.} Hereafter, the optimal linear combination of the standard and conditioned estimators will be employed as a third estimator, as in waste-recycling Monte Carlo~\cite{delmas:2009,adjanor:2011}, because it provides maximum reduction of the statistical variance via a control variate~\cite{caflisch:1998}. 

\blue{\section{Auxiliary absorbing Markov chains}}
\blue{Correlation splitting and conditioning} can be employed in conventional or advanced kMC simulations. Here, we present simulations of severe dynamical trapping using the kinetic path sampling (kPS) algorithm~\cite{athenes:2014,athenes:2019,sharpe:2020}. kPS is an accelerated kMC algorithm developed upon theories of absorbing Markov chains (AMC) and graph transformation~\cite{wales:2009}. It performs non-local displacements avoiding the most trapping states. Those are formally pooled together into a prespecified set, called \emph{transient set} using AMC terminology and denoted by $\mathtt{T}$. Moreover, kPS algorithm is able to generate first-passage paths and times efficiently based on matrix factorizations or inversions. Other AMC-kMC algorithms rely either on matrix diagonalisations~\cite{novotny:1995,opplestrup:2006,oppelstrup:2009,donev:2010} or a mean rate ersatz~\cite{athenes:1997,puchala:2010} based on Eq.~\eqref{eq:time_conditioning}. The kMC and kPS stochastic matrices, respectively denoted by $P^\mathrm{o}$ and $P$, satisfy the conditions  $P^\mathrm{o}_{\chi\chi}=0$ and $P_{\chi\zeta}=0$, $\forall \chi$, $\forall \zeta \in\mathtt{T}\cup\lbrace \chi \rbrace$. The probabilities of not transitioning are always zero from any state and the respective MFPTs, denoted by $\tau^\mathrm{o}_\chi$ and $\tau_\chi$, are state-dependent. Matrix $P$ is constructed from $P^\mathrm{o}$ and an absorbing stochastic matrix, $P^\mathrm{a}$ coinciding with $P^\mathrm{o}$ over $\mathtt{T}$ and with identity matrix $I$ over $\bar{\mathtt{T}}$, the complement of  $\mathtt{T}$. This entails that transitions from $\bar{\mathtt{T}}$ are cancelled, making $\mathtt{T}$-states transient and  $\bar{\mathtt{T}}$-states recurrent: 
\begin{align}
 P^\mathrm{a}_{\chi\chi'} = \bm{1}_\mathtt{T}(\chi) P^{\mathrm{o}}_{\chi\chi'}  + \bm{1}_\mathtt{\bar{T}}(\chi) I_{\chi\chi'}
\end{align}
where indicator function $\bm{1}_\mathtt{S}$ is 1 if $\chi \in \mathtt{S}$ and 0 otherwise.\\ Resorting to AMC theory~\cite{novotny:1995, athenes:1997,wales:2009,opplestrup:2006,oppelstrup:2009,donev:2010,puchala:2010,athenes:2014,athenes:2019,sharpe:2020}, the mean number of visits of transient state $\zeta'$ is defined as the  conditional expectation of the indicator function $\bm{1}_{\lbrace\zeta'\rbrace}$ given initial state $\chi_1 \in\mathtt{T}$. It corresponds to the Green function $\chi_1 \mapsto G^\mathrm{a}_{\chi_1\zeta'}=\sum_{k=0}^\infty \eca{\bm{1}_{\lbrace\zeta'\rbrace}(\chi_{k+1})|\chi_1}$ solution of the Poisson equation over  $\mathtt{T}$:  
\begin{align}\label{eq:poissonGa}
 G^\mathrm{a}_{\chi_1\zeta'} = \eca{G^\mathrm{a}_{\chi_2\zeta'} |\chi_1} + \bm{1}_{\lbrace \zeta' \rbrace}(\chi_1), 
\end{align}
entailing $G^\mathrm{a}_{\zeta\zeta^\prime}= \sum_{\xi\in\mathtt{T}}P^\mathrm{o}_{\zeta\xi}G^\mathrm{a}_{\xi\zeta^\prime} + I_{\zeta\zeta^\prime}$, while $G^\mathrm{a}_{\chi\chi'}$ is set to zero whenever $\chi$ or $\chi'$ lies outside $\mathtt{T}$. Thus, $G^\mathrm{a}$ and $I-P^\mathrm{a}$ are two singular matrices, pseudoinverse of each other, whereas $G^\mathrm{a}_{\zeta\zeta^\prime}$ is the regular inverse of $I_{\zeta\zeta^\prime}-P^\mathrm{a}_{\zeta\zeta^\prime}$ for $\zeta,\zeta^\prime \in \mathtt{T}$. For an absorbing Markov chain starting from $\zeta$, the knowledge of the number of visits of state $\zeta'$ enables one to deduce the probability of being absorbed at $\chi'\in \bar{\mathtt{T}}$ and the mean time for being absorbed in  $\bar{\mathtt{T}}$: 
\begin{align}
\begin{aligned}
\varPi^\mathrm{a}_{\zeta\chi'} &= \sum\nolimits_{\zeta'} G_{\zeta\zeta'}^\mathrm{a}P_{\zeta'\chi'}^\mathrm{a}+I_{\zeta\chi'}, \\
\tau_\zeta^\mathrm{a} &= \sum\nolimits_{\zeta'} G_{\zeta\zeta'}^\mathrm{a} \tau_{\zeta'}^\mathrm{o}. 
\end{aligned} 
\end{align}
The overlying transition probability from $\chi$ to $\chi'\neq\chi$ and the MFPT from $\chi$ eventually write  
\begin{align}
\begin{aligned}
P_{\chi\chi'} &= \sum\nolimits_{\zeta} P^\mathrm{o}_{\chi\zeta}\varPi^\mathrm{a}_{\zeta\chi'}\big/ \pi_\chi 
\\
\tau_\chi &=  \sum\nolimits_{\zeta} P^\mathrm{o}_{\chi\zeta} \left[\tau^\mathrm{o}_\chi + \tau^\mathrm{a}_{\zeta}\right]\big/ \pi_\chi, 
\end{aligned} \label{eq:Pnew}
\end{align}
where rescaling of both transition probabilities and elapsed times by probability $\pi_\chi = 1-\sum_\zeta P^\mathrm{o}_{\chi\zeta}\varPi^\mathrm{a}_{\zeta\chi}$ serves to cancel the flicker probability, thus $P_{\chi\chi} =0$. Besides, if $P^\mathrm{o}_{\chi\chi'}$ obeys detailed balance w.r.t its stationary probability distribution $\rho^\mathrm{o}_\chi$ over $\bar{\mathtt{T}}\cup\mathtt{T}$, then so does $P_{\chi\chi'}$ w.r.t $\rho_\chi \propto \rho^\mathrm{o}_\chi\pi_\chi$ over $\bar{\mathtt{T}}$. The scaled probability distribution $\rho^\mathrm{eq}_\chi\propto \rho^\mathrm{o}_\chi\tau^\mathrm{o}_\chi \propto \rho_\chi\tau^\mathrm{o}_\chi/\pi_\chi$ then corresponds to the thermodynamic equilibrium. Fulfilment of the reversibility property allows \blue{conditioning} to be performed. The conditioned estimator~\eqref{eq:condestimator} can thus be used. \blue{Note that the kPS algorithm reverts to the conventional kMC algorithm when none of the states are made transient.} 


\begin{figure}
\includegraphics[width=0.475\textwidth]{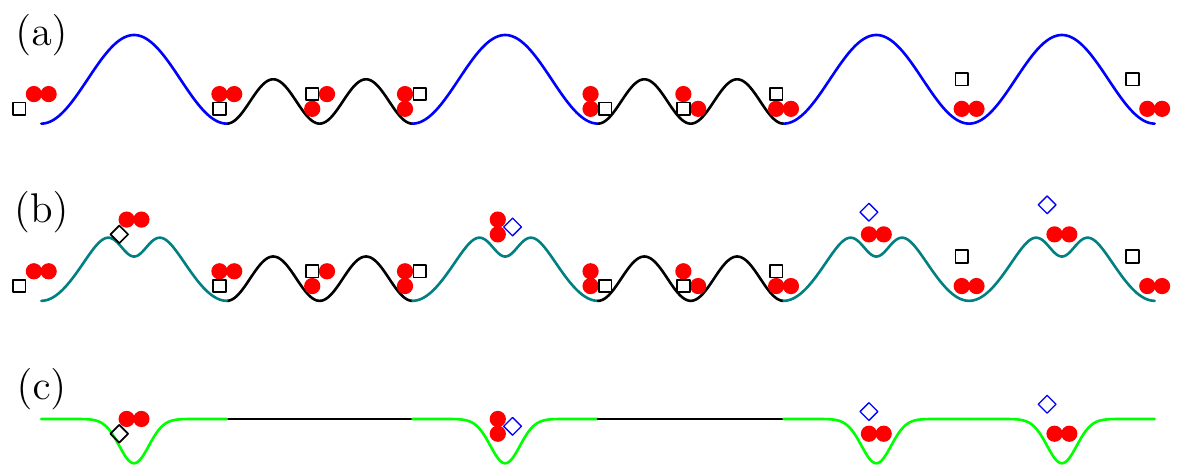}
\caption{Transformation of transition networks: (a) physical network wherein blue and black barriers map out exchanges of a vacancy (empty squares) with bulk $A$-atoms (not displayed) or solute $B$-atoms (red disks); (b) augmented network after inserting saddle states along vacancy-$A$ exchanges (empty diamonds represent vacancies at saddle positions); (c) reduced network after eliminating stable states.}
\label{fig1}
\end{figure}

\blue{\section{Random alloy model}}
We illustrate the approach by computing the mass transport coefficients in a random binary alloy \blue{on a square lattice with periodic conditions. Diffusion is mediated by a single vacancy $V$ exchanging with  $A$ or $B$ nearest-neighbor atoms~\cite{allnatt_lidiard_1993,barbe:2006,trinkle:2018}. Transition rates are usually computed in the framework of classical transition state theory~\cite{hanggi:1990}. They have the form $\nu_X = \nu^\star \exp[-E_X/(k_BT)]$ where $E_X>0$ is the energy barrier that neighbouring atom $X$ jumping into the vacancy must cross, $T$ denotes  temperature, $k_B$ stands for Boltzmann's constant and $\nu^\star$ is the attempt frequency for both species. Here, transition rates are environment-independent and only depend on the type of the jumping atom. Consequently, all states are equiprobable, the system energy has same constant value for all stable states. No thermodynamic transition occurs in the random solid solution~\cite{allnatt_lidiard_1993}. In particular, the site percolation threshold is independent of the jumping frequencies $\nu_A$ and $\nu_B$. Despite its simplicity, the random alloy model is a nontrivial system and is thus customarily used to study trapping and percolation~\cite{barbe:2006,trinkle:2018}. 
Assuming $E_A > E_B > 0$,  the frequency ratio $\nu_B/\nu_A$  increases with decreasing the temperature. Hence, low temperatures result in dynamical trapping. Here, the jump rate of $B$-atoms, $\nu_B $, will be much higher than the one of $A$-atoms, $\nu_A$. For the sake of simplicity, we adopt $\nu_A$ units by setting this value to 1 in the following.}
To perform kPS simulations, the transient graph is to be crafted as the union of many disconnected subgraphs, so that relatively small blocks of $I-P^\mathrm{a}$ need being numerically manipulated on the fly (for computing $\varPi^\mathrm{a}$ and $\tau^\mathrm{a}$). We illustrate the construction of the simulated networks on a square lattice in Fig.~\ref{fig1}. Trapping is suppressed by making transient all the states where the vacancy is located next to a $B$-atom. The difficulty is that vacancy-solute cluster shapes are connected to each others (see Fig.~\ref{fig1}.a), rendering the size of the local transient block so huge that its enumeration is numerically impracticable except for small and isolated $B$-clusters. The problem is mitigated by first augmenting the transition network through the insertion of saddle states along $AV$ exchanges to divide the trapping network into many disconnected subgraphs (see Fig.~\ref{fig1}.b). Transition probabilities from any added saddle state to its two adjacent stable states both equal half. Besides, doubling the transition rates from a stable state to its adjacent saddle states and nullifying the residence times $\tau^{\mathrm{o}}$ at the added saddles compensates the effect of the occasional flickers between stable and saddle states. Finally, stable states are all turned transient, as diagrammed in Fig.~\ref{fig1}.c. That is, set $\mathtt{T}$ is the space of stable states and the simulated network $\bar{\mathtt{T}}$ exclusively consists of the inserted saddle states. The augmentation/reduction of the network leaves the distributions of the sequences of simulated events and times unaffected. 
\begin{figure}
\includegraphics[width=0.475\textwidth]{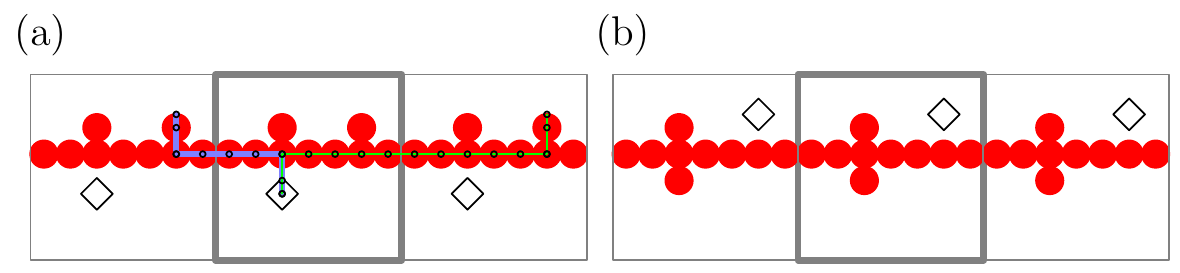}
\caption{\green{A supercell (in bold) is periodically replicated horizontally. The blue and green paths represent two distinct sequences of atomic displacements through the percolating solute cluster (the transient sub-graph) between the same starting state in (a) and ending state in (b).}}
\label{fig2}
\end{figure}

Computing the conditional cumulants given the sampled states in estimator~\eqref{eq:condestimator} is strenuous when \green{a solute cluster, i.e. a sub-graph of the transient network, percolates through the periodically replicated supercell, as illustrated in Fig.~\ref{fig2}}. Cumulants $\m(\chi_0)$ and $\mathbf{V}(\chi_0)$ must account for all the displacements $\del{\x}{0}{1}$ possibly sampled by kPS algorithm. The sampled displacements starting in $\chi_0$ and ending in $\chi_1\in\bar{\mathtt{T}}$ are all congruent to $\dis(\chi_0,\chi_1)$ modulo the cell periods along the percolating directions~\green{(see Fig.~\ref{fig2}.a)}, and many among them may differ from $\dis(\chi_0,\chi_1)$. This consequent sampling burden is avoided by analytically integrating the underlying conditional expected displacements $\m^\mathrm{a}(\chi_h) \triangleq \eca{\x^\mathrm{u}_{h\shortrightarrow h+1}| \chi_h}$, in a similar way to~\cite{athenes:2019,swinburne_perez:2020}. Letting $\mathbb{E}^\mathrm{o}\left[\cdot\right]$ denote expectation  w.r.t. $P^\mathrm{o}$, both cumulants of overlying displacement $\x_{0\shortrightarrow 1}$ are determined by the underlying sequences $\x^\mathrm{u}_{0\shortrightarrow 1}+\x^\mathrm{u}_{1\shortrightarrow \infty} \equiv \x_{0\shortrightarrow 1}$ \green{associated with the successive states of the AMC, $\lbrace\chi_{h}^\mathrm{u}\rbrace_{h \geq 1}$, and where $\chi_0^\mathrm{u}=\chi_0$ and $\chi_\infty^\mathrm{u}=\chi_1$.}

Conditioning after the $P^\mathrm{o}$-transition within decomposition~\eqref{eq:Pnew} yields
\begin{subequations}
\begin{align}
\m(\chi_0) & = \eco {\x^\mathrm{u}_{0\shortrightarrow 1} + \eca {\x^\mathrm{u}_{1\shortrightarrow \infty}|\chi_1}|\chi_0}/\pi_{\chi_{0}}, \\
\mathbf{V}(\chi_0) &= \eco{\left[\x^\mathrm{u}_{0\shortrightarrow 1}+\eca{\x^\mathrm{u}_{1\shortrightarrow \infty} |\chi_1}- \m(\chi_0)\right]^2\big|\chi_0} /\pi_{\chi_0}. 
\end{align}\label{eq:conditioned_cum}
\end{subequations}
The mean displacement $\eca{\m^\mathrm{a}}$ being null, the relaxation vector $\bmu^\mathrm{a}(\chi_1) \triangleq \eca{\delx{1}{\infty} |\chi_1}=\sum\nolimits_{h=1}^{+\infty}\eca{\m^\mathrm{a}(\chi_{h})| \chi_1}$, is null outside $\mathtt{T}$ and satisfies a Poisson equation inside $\mathtt{T}$
\begin{align}
 \bmu^\mathrm{a}(\chi_1)  = \eca{ \bmu^\mathrm{a}(\chi_2)|\chi_1} + \m^\mathrm{a}(\chi_1).   
\end{align}
The solution is unique, reads $ \bmu^\mathrm{a}(\zeta) = \sum_{\zeta'} G^\mathrm{a}_{\zeta\zeta'} \m^\mathrm{a}(\zeta')$ for $\zeta \in \mathtt{T}$ and cancels outside ${\mathtt{T}}$. It serves to evaluate the two conditional cumulants~\eqref{eq:conditioned_cum} \green{for estimator~\eqref{eq:condestimator}. We resort} to their \green{respective} LTE and LTV expressions with  conditioning over the underlying absorbing chain initiated from $\zeta\equiv \chi_1$ \green{to obtain a numerically tractable expression }
\begin{subequations} 
\begin{align}
\m(\chi) &= \sum\nolimits_{\zeta}  \left[\dis(\chi,\zeta) + \bmu^\mathrm{a}(\zeta)\right]{P^\mathrm{o}_{\chi\zeta} }/{\pi_{\chi}}, \\
\mathbf{V}(\chi)
&= \sum\nolimits_{\zeta} \hspace{-1mm}\left\{ \hspace{-0.5mm}\mathbf{V}^\mathrm{a}(\zeta) +\left[\dis(\chi,\zeta)+\bmu^\mathrm{a}(\zeta)-\m(\chi) \right]^2 \hspace{-0.25mm} \right\} \hspace{-0.5mm}{P^\mathrm{o}_{\chi\zeta} }/\pi_{\chi}, 
\end{align}
\end{subequations}
where $\chi \equiv \chi_0$ in~\eqref{eq:conditioned_cum}. 
The variance of the absorbing chain conditioned on $\zeta\equiv \chi_1$ is also computed using the Green function and relaxation vectors, resorting to relation~\eqref{eq:lastofderivation} below: 
\begin{align}
\mathbf{V}^\mathrm{a}(\chi_1) &= \eca{ \left.\left(\sum\nolimits_{h=1}^\infty \x^\mathrm{u}_{h\shortrightarrow h+1}\right)^{2} \right|\chi_1 }-\bmu^\mathrm{a}(\chi_1)^{2}  \\ 
& \hspace{-8mm}=\sum\nolimits_{h=1}^\infty \eca{\left.(\x^\mathrm{u}_{h \shortrightarrow h+1}+\bmu^\mathrm{a}(\chi_{h+1}))^{2} - \bmu^\mathrm{a}(\chi_{h})^{2} \right|\chi_1} \nonumber
\\ 
&\hspace{-8mm}= \sum\nolimits_{\chi \zeta} G^\mathrm{a}_{\chi_1\chi} \left[P^\mathrm{a}_{\chi\zeta}\left[ \dis(\chi,\zeta)+ \bmu^\mathrm{a}(\zeta) \right]^{2}-\bmu^\mathrm{a}(\chi)^{2}\right]. \label{eq:lastofderivation}
\end{align}
\green{Note that q}uantities $\bmu^\mathrm{a}$ and $\mathbf{V}^\mathrm{a}$ are first and second derivatives of a displacement cumulant generating function at its origin, respectively~\cite{swinburne_perez:2020}.  \\

\blue{\section{Measurements of diffusion coefficients}}

We perform a series of kPS simulations of extreme dynamical trapping by setting $\nu_B =10^5$ and $\nu_A=1$ in a periodically replicated square lattice of size $16\times 16$. Compositions are gradually increased from 1 to $254$ $B$-atoms. The atomic concentration of $X$ is denoted by $C_X$, entailing that $C_{A+B}=C_A+C_B$ is one. The diffusion matrix  $\mathbb{D}(\dis)$ is estimated using $\mathbf{D}_{l,L}^\mathrm{std}$ and  $\mathbf{D}_{l,L}^\mathrm{cnd}$ estimators with $L=\ell_\mathrm{max}=200$,  $l=5\cdot 10^5$ for $C_B<0.5$ and $l=10^4$ for $C_B\ge 0.5$.  Estimates $D_X^\mathrm{std}$ and $D_X^\mathrm{cnd}$ of the diffusion coefficients $D_X$ for both species and the vacancy ($X=A$, $B$ and $A+B$) are then obtained via $\mathbf{D}_{l,L}^\mathrm{std}$ and  $\mathbf{D}_{l,L}^\mathrm{cnd}$ by averaging the corresponding elements over the two space directions. The optimal estimate is then obtained as $D_X^\mathrm{opt}=(1-c^\star)D_X^\mathrm{std} + c^\star D_X^\mathrm{cnd}$ where the optimal control variate is $c^\star = -\mathrm{cov}\left(D_X^\mathrm{std},D_X^\mathrm{cnd}-D_X^\mathrm{std}\right)/\mathrm{var}\left(D_X^\mathrm{cnd} -D_X^\mathrm{std}\right) $. We furthermore record the transient regimes $ D^\mathrm{cnd}_X(\ell) = \varSigma_\ell +\frac{1}{\ell} \varGamma_1 +\frac{1-\ell}{\ell}\varGamma_{\ell-1}$ from~\eqref{eq:LTC} where $\varSigma_\ell$ and $\varGamma_{\ell-1}$ are estimates of the corresponding intra- and extra-correlated parts. Last, we define and evaluate the intra-to-extra correlation ratio $\phi=\varSigma_L/\varGamma_L$ and the reduction factor of the statistical variance $\eta = \mathrm{var}\left(D^{\mathrm{std}}_X \right)/\mathrm{var}\left(D^{\mathrm{opt}}_X \right)$. 

\begin{figure}
\includegraphics[trim= 4mm 2mm -3mm 0.0mm, clip,width=0.5\textwidth]{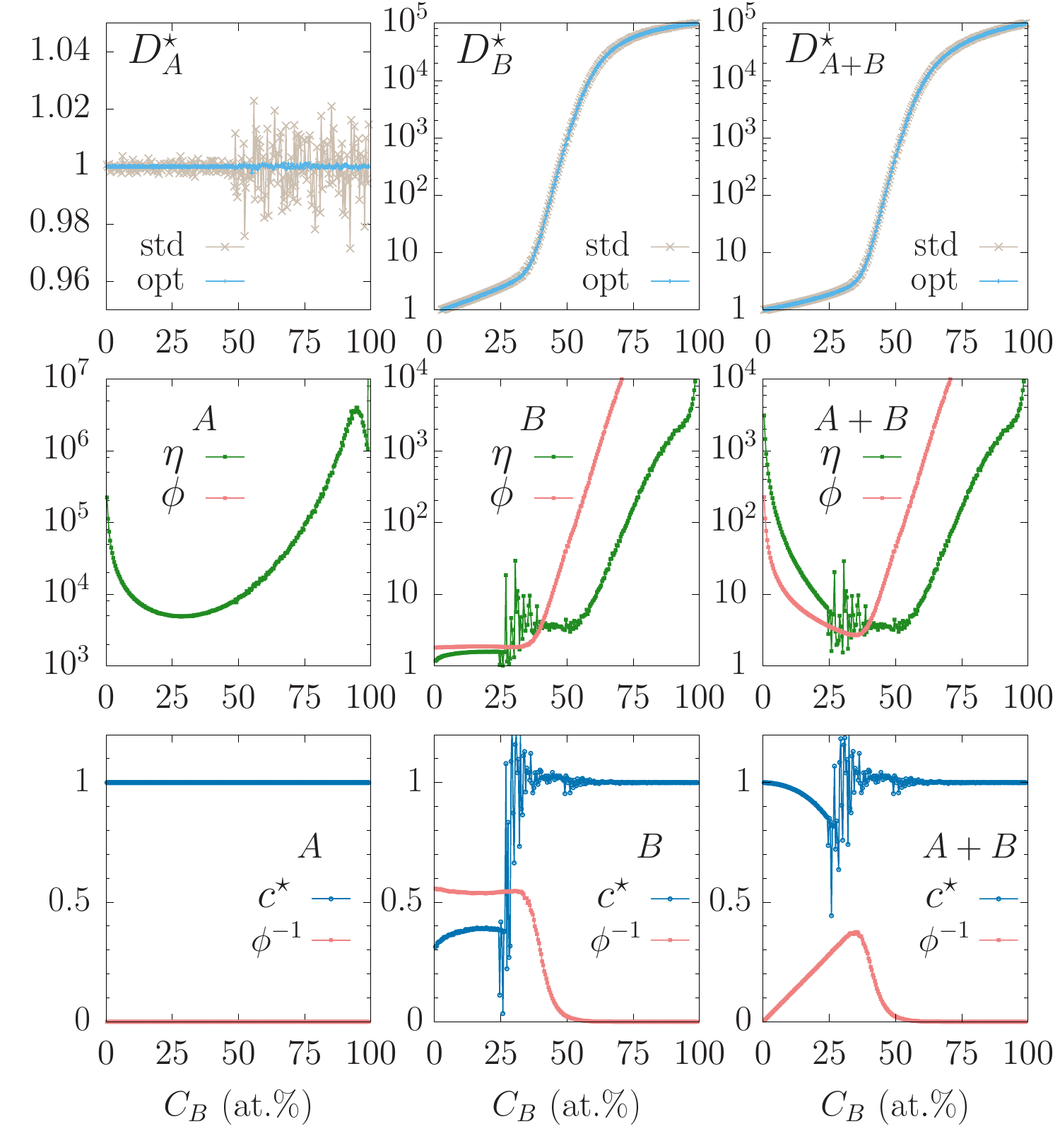}
\caption{Atomic diffusivities $D^\star_X$ ($X=A$, $B$ and $A+B$) for extreme dynamical trapping ($\nu_A=1$ and $\nu_B=10^5$), as a function of $C_B$, with corresponding variance reduction factor $\eta$, intra-to-extra correlation ratio $\phi$ and optimal variate $c^\star$. The supercell size is $16\times16$.}
\label{fig3}
\end{figure}

\begin{figure}
\includegraphics[trim= 5mm 12mm 0 0, clip ,width=0.5\textwidth]{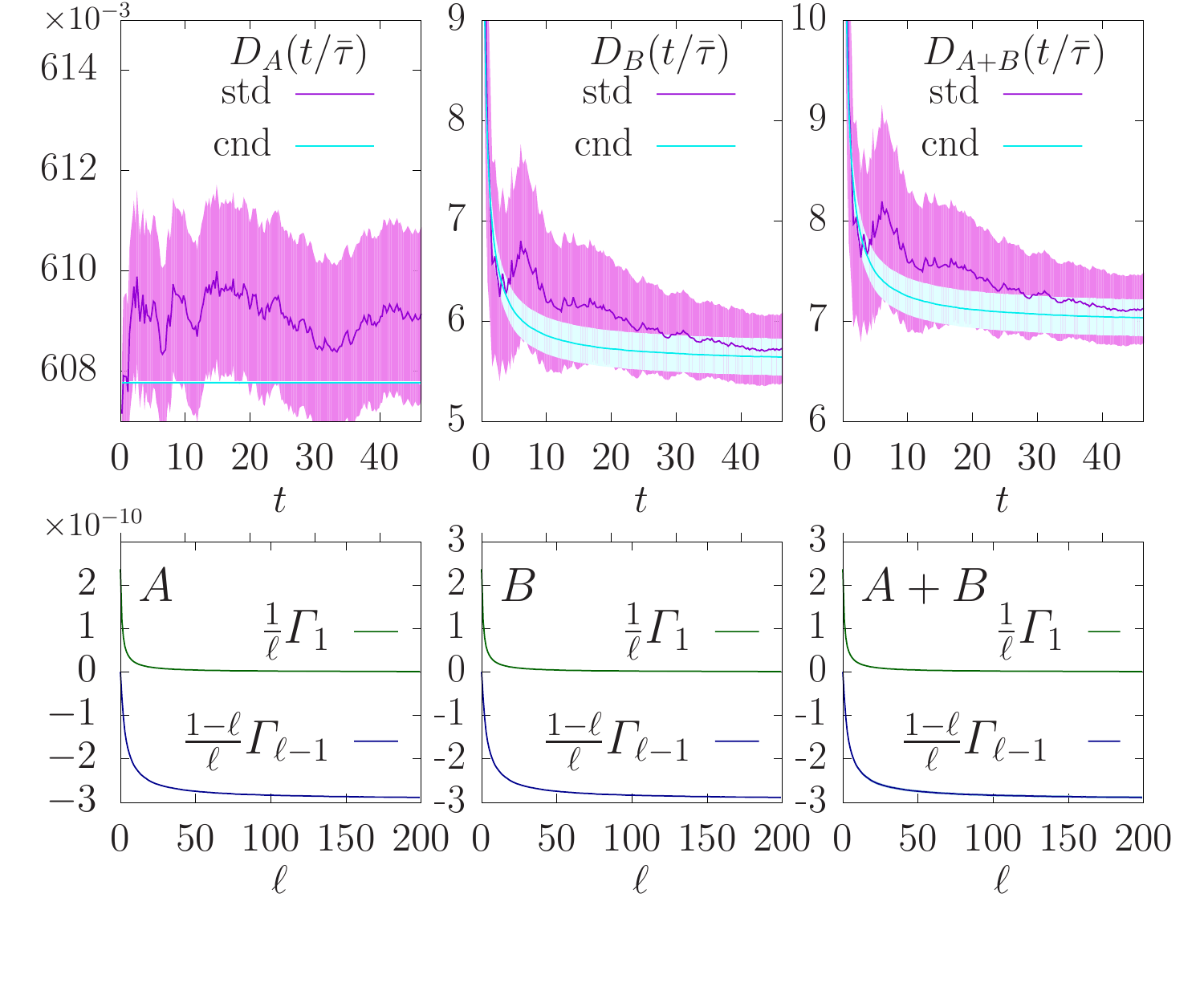}
\caption{Time dependence of the estimated $D_X(\ell)$ with $\ell = t/\bar{\tau}$, using standard~\eqref{eq:stdestimator} and conditioned~\eqref{eq:condestimator} estimators, for $X=A$, $B$ and $A+B$. Filled areas around the curves represent $95\%$ confidence intervals (CI). \green{CIs around $\tfrac{1}{\ell}\varGamma_1$ and $ \tfrac{1-\ell}{\ell} \varGamma_\ell$ curves are too small to be visible.}}
\label{fig4}
\end{figure}

The atomic diffusivities $D^\star_X=D_X/C_X$ evaluated using the standard and optimal estimators for the entire composition range are displayed in Fig.~\ref{fig3}, together with $\eta$, $\phi$ and $c^\star$. 
We observe that the speedup measured in term of $\eta$ is considerable whenever the intra-correlated contribution to diffusivity dominates ($\phi >> 1$). In this situation, the optimal estimator perfectly matches the conditioned estimator ($c^\star=1$).  The optimal estimator is relatively less efficient for the fast-diffusing $B$-atoms when the extra-correlated contribution to $B$-diffusivity is significant, at $B$-concentration lower than 40\%. The lowest reductions of variance are about a factor of 2. To investigate the origin of the relatively lower performance of the conditioned estimator, we display in Fig.~\ref{fig4} the transient regime estimated from $D^\mathrm{std}_X(\ell)$, $D^\mathrm{cnd}_X(\ell)$ and $\varGamma_{\ell-1}$ with estimates of their standard errors for $0\leq \ell \leq \ell_\mathrm{max}$. The errors associated with $\varGamma_{\ell-1}$ terms are negligible compared to those with $\varSigma_\ell$ terms and are attributed to the intermittent binding of the vacancy to the $B$-clusters. Removing this intermittency in the kPS algorithm would require making additional states transient and handling $A$-connected subgraphs similarly to $B$-clusters. 

\blue{Interestingly, the negative contributions $\tfrac{1-\ell}{\ell}\varGamma_\ell$ in Fig.~\ref{fig4} undergo fast algebraic decays towards their plateau values $\delta$. The three curves are very well fitted by a law of the form $g(\ell+\ell_0)^\alpha + \delta$. Implementing the nonlinear least-squares Marquardt-Levenberg algorithm with the four fitting parameters $g$, $\ell_0$, $\alpha$ and $\delta$ yields the power-law exponents $\alpha$ associated with $A$, $B$ and $A+B$ diffusivities. The obtained $\alpha$ values lie in the following $68\%$-confidence intervals: $-1.302\pm 0.091$, $-1.015\pm 0.008 $ and $-1.015 \pm  0.008$, respectively. The convergence features of the mean-squared mean-displacements entails that a standard diffusive regime exhibiting short-range dependence is quickly reached as $\ell$ increases. This property is related to the fast algebraic decay of the autocorrelation function, as detailed in~\cite{noteSM}. Note that the larger exponent associated with $A$-diffusivity decreases as the frequency ratio is decreased~\cite{noteSM}. This physical trend results from a nontrivial interplay between composition, transition rates, site interactions and geometry. We refer to the recent work of Settino \emph{et al.}~\cite{settino:2020} who studied anomalous diffusion in the Aubry-André model via the  autocorrelation functions and investigated the dependence of their power-law decay on the model physical parameters.}

\begin{figure}[h!]
\includegraphics[trim= 0mm 0mm 0 0, clip ,width=0.45\textwidth]{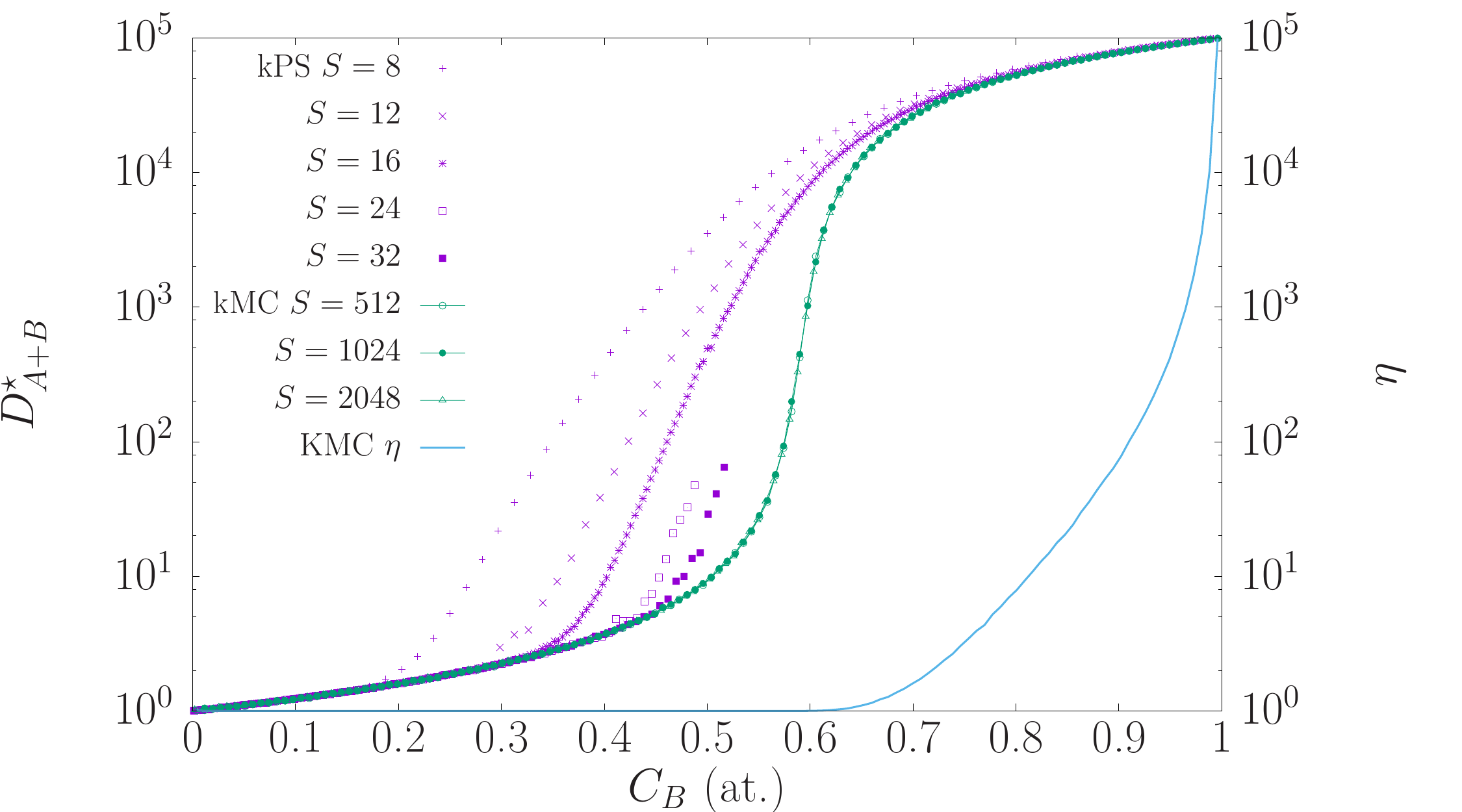}
\caption{\blue{Diffusion coefficient $D_{A+B}$ as a function of composition for various lattice sizes S using the kPS and kMC algorithms. The optimised estimator is used in all measurements. The kPS simulations are performed with same set-up as in Fig.~\ref{fig3}, but four additional supercell sizes, $S=8$, 12, 24 and 32. The conventional kMC simulations employ $10^4$ trajectories of $10^6$ steps. Their CPU costs are almost independent of the system sizes. Note that the speed-ups $\eta$ obtained through conditioning are considerable in conventional kMC simulations of dilute $A_{x}B_{1-x}$ systems with $x<0.1$.}}
\label{fig5}
\end{figure}

Here, the extremely severe dynamical trapping of the vacancy with $B$-atoms strongly impacts the monotonous increase of $B$-diffusivity with $C_B$. The sudden sharp increase of $D_B^\star$ observed at $C_B \approx 0.35$ occurs at a concentration much lower than $p_s=0.592745$, the site percolation threshold~\cite{ziff:1992}. This dynamical transition is due to the cooperative motion of small $B$-clusters whose effectiveness increases considerably with $C_B$ \green{in small supercells. With increasing supercell sizes, the transition further sharpens and occurs closer to the percolation threshold, as shown in Fig.~\ref{fig5}}.

\blue{The kPS simulations for the two largest supercell sizes in Fig.~\ref{fig5} became inefficient and were stopped before terminating. The inefficiency results from the larger sizes of the percolating solute clusters and to the cost of computing the Green functions for the associated transient sub-graphs, a task involving matrix inversion or factorisation. Here, a direct non-optimized dense solver was used whose complexity scales cubically with the sizes of the transient sub-graphs. As a result, the approach starts consuming CPU time enormously when supercell sizes and $B$-concentrations are too large. 
Nevertheless, dynamical trapping disappearing beyond the percolation threshold, conventional kMC simulations can be used instead. One then obtains reliable estimates of the diffusion coefficients at much larger system sizes, as shown in Fig.~\ref{fig5}}. 
\blue{The variance reduction factor $\eta$ allowed by the optimal estimator in kMC simulations increases from 1 below the percolation threshold to $10^5$ in pure $B$. For concentrated alloy system $A_{0.1}B_{0.9}$, the factor of variance reduction $\eta$ is $10^2$, yielding an appreciable speed-up.}
\blue{However, the conditioned estimator underperforms the standard one at the low-$B$ concentrations, when trapping becomes important in conventional kMC simulations. Results obtained using the conventional kMC algorithm in systems exhibiting moderate dynamical trapping are also reported in~\cite{noteSM}, where it is shown that significant variance reductions via conditioning are systematically observed.} 
\blue{The conventional kMC algorithm outperforms kPS algorithm with excessively large transient sub-graphs because the trapping strength measured in number of steps the vacancy remains attached to a solute cluster scales linearly with the cluster size, while the factorization cost scales cubically here. A smaller pre-factor for the factorization costs makes kPS advantageous for transient sub-graphs smaller than a certain threshold value. Hence, there are two complementary ways to improve kPS simulations: (i) avoid turning a large vacancy-solute cluster into a transient sub-graph when its size is above the threshold, (ii) resort to a sparse linear solver to reduce the factorization/inversion costs. The latter costs grow quadratically with the sub-graph sizes for the direct multi-frontal linear solver used in Ref.~\cite{athenes:2019}.}

\blue{Nonetheless, an interesting trend emerges from the results of the kPS simulations with severe trapping. We indeed observe that the extra-to-intra correlation ratio $\phi^{-1}$ vanishes for all diffusion measurements reported in Fig.~\ref{fig3} when the solute concentrations are larger than the dynamical percolation threshold values. This property means that the correlations of the underlying process are well captured by the auxiliary absorbing Markov chains. It entails that the extra-correlated part of the diffusion matrix is negligible and justifies the use of the kinetic cluster expansion (KineCluE) method~\cite{schuler:2020} to solve this particularly difficult diffusion problem. This method neglects the extra-correlated part and similarly resorts to auxiliary absorbing Markov chains, but these ones are constructed from a sub-graph expansion. To adapt the KineCluE method to concentrated alloys, the absorbing Markov chains should be solved for a sample of alloy configurations generated at thermodynamic equilibrium. Whenever the extra-correlated part containing the cluster correlations is negligible, the transport coefficients may be correctly estimated from their intra-correlated contributions exclusively.}

\blue{Concerning the transport properties of the slowly migrating solvent species,} we observe that $A$-diffusivity is almost not impacted by dynamical trapping and remains constant. Reproducing a sluggish diffusion behaviour where $A$-diffusivity exhibits a minimum at intermediate compositions, as observed experimentally in several concentrated solid solution alloys~\cite{athenes:1997} and highly debated in the literature on high entropy alloys~\cite{tsai:2013,paul:2017,miracle:2017,osetsky:2018,kottke:2020,daw:2021}, will necessitate to \green{parameterize the exchange rates $\nu_A$ and $\nu_B$}, for instance by accounting for the binding energies and entropies between alloying elements and vacancies~\cite{trochet:2017,trochet:2019,swinburne:2018,lapointe:2020}. This would affect short-range ordering or clustering, introduce a dependency of vacancy concentration on alloy composition and impact atomic diffusion.

\blue{\section{Conclusion}}

\blue{Splitting the diffusion matrix into its intra- and extra-correlated parts and conditioning the correlations enables one to formulate a law of total diffusion.} A conditioned estimator then allows measuring atomic diffusivities with considerably improved accuracy \green{in absence of dynamical trapping}. The use of an optimized control variate guarantees systematic variance reductions compared with standard computations performed with conventional or advanced kinetic Monte Carlo methods. Conditioning is particularly useful when the latter methods employ auxiliary absorbing Markov chains able to suppress trapping at the underlying scale, like kinetic path sampling, as it leverages the Green functions that are computed to generate the escaping first-passage paths. 
\blue{However, the cost of computing these Green functions must remain reasonable so that kinetic path sampling simulations always outperforms conventional kinetic Monte Carlo simulations. If this condition is not met, it is preferable to limit the sizes of the transient sub-graphs or to revert to conventional kMC simulations.}  
\blue{The conditioning approach is in principle amenable to the numerous kMC simulations dealing with realistic potential energy surface accounting for lattice distortions~\cite{trochet:2019} or describing amorphous structures~\cite{li:2020}. In these simulations, the computation of the transition rates is often so costly that the number of simulated events is quite limited. The variance reduction technique proposed here will certainly be useful in this particular context. As the conditioning approach represents a small overhead, it can be readily applied to measure transport properties of any particles migrating through thermally activated events in crystalline or non-crystalline solids. Its range of applicability encompasses surface and bulk diffusion in presence of extended defects~\cite{grabowski:2018,athenes:2019,landeiro:2020,rahman:2021} like cavities, clusters, dislocations or grain boundaries, within complex alloys or any material compounds. 
}
\begin{acknowledgments}
Stimulating discussions with Vasily Bulatov, Benjamin Jourdain, Maylise Nastar, Kirone Mallick and Dallas Trinkle are gratefully acknowledged.   
\end{acknowledgments}


\end{document}